\documentclass[12pt]{article}
\usepackage{amstex,amssymb}
\def\1#1{{\bf #1}}
\def\2#1{{\cal #1}}\def\9#1{{\sl #1}}\def\4#1{{\tt #1}}\def\5#1{{\sf #1}}
\def\6#1{{\mathfrak #1}}\def\7#1{{\mathbb #1}}\def\8#1{{\rm #1}}
\def\9#1{{\mit #1}}

\def\hom{{\rm Hom}}
\def\aut{{\rm Aut}}
\def\3{{\ss}}

\def\skb{\vskip 0.5cm}

\def\beq{\begin{eqnarray}}
\def\eeq{\end{eqnarray}}
\def\vs{\vspace{0.2cm} \\}

\def\Poin{{\8P^\uparrow_+}}
\def\Lor{{\8L^\uparrow_+}}
\newtheorem{The}{Theorem}[section]
\newtheorem{Def}[The]{Definiton}
\newtheorem{Lem}[The]{Lemma}
\newtheorem{Pro}[The]{Proposition}
\newtheorem{Cor}[The]{Corollary}
\def\bdef{\begin{Def}\1: \em}
\def\eef{\end{Def}}

\def\blem{\begin{Lem}\1: }
\def\elem{\end{Lem}}
\def\bthe{\begin{The}\1: }
\def\ethe{\end{The}}
\def\bpro{\begin{Pro}\1: }
\def\epro{\end{Pro}}
\def\bcor{\begin{Cor}\1: }
\def\ecor{\end{Cor}}

\def\supp{{\rm supp}}

\def\tr{{\rm tr}}
\def\al{\alpha}
\def\be{\beta}
\def\gam{\gamma}\def\Gam{\Gamma}
\def\lam{\lambda}
\def\eps{\epsilon} 
\def\te{\theta}

\def\Sgm{\Sigma}
\def\om{\omega}\def\Om{\Omega}

\def\bpr{\paragraph*{\it Proof.}}

\def\epr{$\Box$\skb}

\def\<{\langle}
\def\>{\rangle}
\def\Ad{{\rm Ad}}

\def\bdes{\begin{enumerate}}
\def\edes{\end{enumerate}}
\newcommand\itno[1]{\item[{\it ({#1})}]}

\def\bmat{\left( \begin{array}{ccc} }
\def\emat{\end{array} \right)}

\def\beqa{\begin{eqnarray*}}
\def\eeqa{\end{eqnarray*}}
\def\bdia{\begin{diagram}}
\def\edia{\end{diagram}}

\def\ul{\underline}

\def\rTo{ \ \longrightarrow \ }
\def\rMapsto{ \ \longmapsto \ }

\title{\bf From euclidean field theory to quantum field theory}
\author{{\it Dirk Schlingemann} \\
II. Institut f\"ur Theoretische Physik \\
Universit\"at Hamburg \\ and \\
The Erwin Schr\"odinger International Institute \\ 
for Mathematical Physics (ESI)\\
Vienna}
\date{}
\begin{document}
\maketitle
\abstract{In order to construct examples for 
interacting quantum field theory models, the methods of euclidean  
field theory turned out to be powerful tools since they 
make use of the techniques of classical statistical mechanics. 

Starting from an appropriate set of 
euclidean $n$-point functions (Schwinger distributions),
a Wightman theory can be reconstructed by 
an application of the famous Osterwalder-Schrader reconstruction theorem.
This procedure (Wick rotation), which  
relates classical statistical mechanics and  
quantum field theory, is, however, somewhat subtle. 
It relies on the analytic properties of the 
euclidean $n$-point functions.

We shall present here a C*-algebraic version of 
the Osterwalder-Scharader reconstruction theorem. We shall see that, 
via our reconstruction scheme, 
a Haag-Kastler net of {\em bounded} operators can directly 
be reconstructed. 

Our considerations also include 
objects, like Wilson loop variables, which are not 
point-like localized objects like distributions. 
This point of view may also be helpful for 
constructing gauge theories.}
\newpage

\section{Introduction}

\paragraph{Why euclidean field theory?}
During the last two decades it turned out that 
the techniques of euclidean field theory are  
powerful tools in order to construct 
quantum field theory models. 
Compared to the method of canonical quantization in Minkowski space,
which, for example, has been used for the construction of 
$P(\phi)_2$ and Yukawa$_2$ models
\cite{GlJa1,GlJa2,GlJa3,GlJa4,Schra0,Schra1}, the functional integral 
methods of euclidean field theory simplify the construction of 
interactive quantum field theory models. 

In particular, 
the existence of the $\phi^4_3$ model as a Wightman theory 
has been established by using euclidean methods 
\cite{FeldOst,SeilSim76,MagSen76} 
combined with the famous Osterwalder-Schrader 
reconstruction theorem \cite{OstSchra1}.    
For this model the methods of canonical quantization are much more 
difficult to handle and lead by no means so far as euclidean techniques do. 
Only the proof of the positivity of the 
energy has been carried out within the hamiltonian framework 
\cite{GlJa1,GlJa5}. 
   
One reason why the functional integral point of view 
simplifies a lot is that the theory of  
classical statistical mechanics can be used.  
For example, renormalization group analysis \cite{GawKup} and 
cluster expansions \cite{Br} can be applied 
in order to perform the continuum and the infinite volume limit of a 
lattice regularized model. 
Instead of working with non-commutative objects, 
one considers the moments 
\beqa
\6S_n(x_1,\cdots,x_n)=\int\8d\mu(\phi) \ \phi(x_1)\cdots\phi(x_n)
\eeqa
of reflexion positive measures
$\mu$, usually called 
Schwinger distributions or euclidean correlation functions, 
on the space of tempered distributions. 

Heuristically, the functional integral point of view leads to 
conceptionally simple construction scheme for a quantum field theory. 
Starting from a given lagrangian density $L$, 
the measure $\mu$ under consideration is simply given by 
\beqa
\8d\mu(\phi)&=&Z^{-1}\ \bigotimes_{x\in\7R^d}\8d\phi(x) \ 
\exp\biggl(-\int \8dx \ L(\phi(x),\8d\phi(x))\biggr)
\eeqa
where the factor $Z^{-1}$ is for normalization.  Therefore, the 
lagrangian $L$ can be interpreted as a germ of a 
quantum field theory. Moreover, this also leads 
to a nice explanation of the minimal action principle.
However, to give the expression above a rigorous mathematical 
meaning is always accompanied with serious technical difficulties.

\paragraph{Some comments on the Osterwalder-Schrader reconstruction
theorem.}
In order to motivate the main purpose of our paper, we shall 
make some brief remarks on the Osterwalder-Schrader reconstruction
theorem \cite{OstSchra1} which relates Schwinger and  Wightman distributions.
Let $T(S)$ be the tensor algebra over the 
space of test functions $S$ (in $\7R^d$) and let us denote by 
$J_E$ ($E$ stands for {\em euclidean}) the two-sided 
ideal in $T(S)$, which is generated by elements 
$f_1\otimes f_2-f_2\otimes f_1\in T(S)$ where  
$f_1$ and $f_2$ have disjoint supports. We build the 
algebra $T_E(S):=T(S)/J_E$ and take the closure $T^\2T_E(S)$ of 
it in an {\em appropriate} locally convex topology. We claim that 
the euclidean group $\8E(d)$ acts naturally by automorphisms 
$(\al_g,g\in\8E(d))$ on $T^\2T_E(S)$.

A linear functional $\eta\in T^\2T_E(S)^*$ fulfills 
the Osterwalder-Schrader axioms if the following conditions hold:
\begin{description}
\itno{E0} $\eta$ is continuous and unit preserving: $<\eta,\11>=1$.
\itno{E1} $\eta$ is invariant under euclidean transformations:
$\om\circ\al_g=\om$. 
\itno{E2} $\eta$ is reflexion positive:
The sesqui-linear form $a\otimes b\mapsto <\eta,\iota_e(a^*)b>$ is 
a positive semi-definite   
on those elements which are localized at 
positive times with respect to the direction $e\in S^{d-1}$
where $\iota_e$ is the automorphism which corresponds to the 
reflexion $e\mapsto -e$. 
\end{description}

Given a linear functional $\eta$ which satisfies the conditions 
{\it (E0)} to {\it (E2)}, the analytic properties 
of the distributions 
\beqa
\6S_n(f_1,\cdots ,f_n)&:=&<\eta,f_1\otimes\cdots\otimes f_n>
\vs\vs
\mbox{ and } \ \ 
S_n(\xi_1,\cdots,\xi_n)&=& \6S_{n+1}(x_0,\cdots,x_n) \ \ ; \ 
\ \xi_j=x_{j+1}-x_j
\eeqa
lead to the result:

\bthe
There exists a distribution $\tilde W_n\in S'(\7R^{nd})$ supported in 
the n-fold closed forward light cone $(\bar V_+)^n$ 
which is related to $S_n$ by the Fourier-Laplace transform:
\beqa
S_n(\xi)=\int \8d^{nd}q \ \exp(-\xi^0q^0-\8i\vec{\xi}\vec{q}) \ \tilde W_n(q)
\eeqa
\ethe

The proof of this Theorem \cite{OstSchra1} relies essentially on the  
choice of the topology $\2T$.
It does not apply for the ordinary $S$-topology, i.e. 
it is not enough to require that the $\6S_n$'s are tempered distributions.  
This was stated wrongly in  
the first paper of \cite{OstSchra1} and was later 
corrected in the second one. We claim that, nevertheless,  
the Theorem might be true for the ordinary $S$-topology, but, at the 
moment, there is no correct proof for it.   
These problems show that the relation 
between euclidean field theory and quantum field theory
is indeed subtle.

In order to formulate the famous Osterwalder-Schrader reconstruction theorem
from a more algebraic point of view,
we shall briefly introduce the notion of a 
{\em local net} and a {\em vacuum state}. 

\subparagraph{\it $\Poin$-covariant local nets:} 
A $\Poin$-covariant local net of *-algebras is an 
isotonous \footnote{Isotony: 
$\2O_1\subset \2O_2$ implies $A(\2O_1)\subset A(\2O_2)$.}   prescription 
$\ul A:\2O\mapsto A(\2O)$, which assigns to each double cone 
$\2O=V_++x\cap V_-+y$ a unital *-algebra $A(\2O)$, on which the  
the Poincar\' e group $\Poin$ 
acts covariantly on $\ul A$, i.e. there is a 
group homomorphism $\al\in\hom(\Poin,\aut A)$, such that 
$\al_gA(\2O)=A(g\2O)$. Here $A$ denotes the *-inductive limit of the net 
$\ul A$. Furthermore, 
the net fulfills locality, i.e. if $\2O,\2O_1$ are two
space-like separated regions $\2O\subset\2O_1'$ then 
$[A(\2O),A(\2O_1)]=\{0\}$. 
A $\Poin$-covariant local net of C*-algebras is called
a {\em Haag-Kastler} net.

\subparagraph{\it Vacuum states:}
A state $\om$ on $A$ is called a vacuum state iff 
$\om$ is $\Poin$-invariant
(or translationally invariant), i.e. $\om\circ\al_g=\om$ 
for each $g\in\Poin$, and for each $a,b\in A$
\beqa
\int \8dx \ <\om,a\al_{(1,x)}(b)> \ f(x)  &=& 0 
\eeqa
for each test function $f\in S$ with 
$\supp(\tilde f)\cap \bar V_+=\emptyset$.   
This implies that there exists a strongly continuous 
representation $U$ of $\Poin$ 
on the GNS Hilbert space of $\om$ such that 
\beqa
U(g)\pi(a)U(g)^*=\pi(\al_ga)
\eeqa
and the spectrum of $U(1,x)$ is contained in the closed forward light cone.  
Here $\pi$ is the GNS representation of $\om$.

Usually it is   required that a vacuum state $\om$ is a 
pure state. This aspect is not so important for our purpose and 
we do not assume this here.
\skb 

An example for a $\Poin$-covariant local net of *-algebras is given by 
the prescription 
\beqa
{\ul T}_M(S):\2O\rMapsto T_M(S(\2O))
\eeqa
where $T_M(S):=T(S)/J_M$\footnote{The ideal 
$J_M$ ($M$ stands for Minkowski) is the two-sided 
ideal in $T(S)$, which is generated by elements 
$f_1\otimes f_2-f_2\otimes f_1\in T(S)$ where  
$f_1$ and $f_2$ have space-like separated supports.} 
is the well known {\it Borchers-Uhlmann algebra}.
We should mention here that now the test functions in $S$ are 
test functions in {\em Minkowski space-time}. 
 
Let $\tau\in \hom(\Poin,\8{GL}(S))$ be the action of 
the Poincar\' e group on the test functions which is given by 
$\tau_gf=f\circ g^{-1}$ then 
\beqa
\al_g(f_1\otimes\cdots\otimes f_n)&:=& \tau_gf_1\otimes\cdots\otimes \tau_gf_n 
\eeqa
defines a covariant action of $\Poin$ on ${\ul T}_M(S)$. 
Now, the theorem above leads to the famous 
Osterwalder-Schrader reconstruction theorem:

\bthe
Given a linear functional $\eta$ which satisfies the conditions 
{\it (E0)} to {\it (E2)}, then there exists a vacuum 
state $\om_\eta$ on the Borchers algebra $T_M(S)$
such that 
\beqa
<\om_\eta,f_1\otimes\cdots\otimes f_n>&=&\6W_n(f_1,\cdots ,f_n)
\eeqa
where $\6W_n$ is defined by
\beqa
\6W_n(x)=\int \8d^{nd}q \ \exp(-\8i\xi q) \ \tilde W_n(q)
&;&  \xi_j=x_{j+1}-x_j \ \ .
\eeqa
\ethe

The fact that $\om_\eta$  is a vacuum state 
on the Borchers algebra is completely equivalent to the 
statement that the distributions $\6W_n$ fulfill the 
Wightman axioms  
in its usual form (except the clustering)(see \cite{StrWgh89}). 

\paragraph{A heuristic proposal for the 
treatment of gauge theories.}
As mentioned above, the main reason for using euclidean field 
theory is for constructing quantum field theory models 
with interaction. In four space time dimensions, the 
most promising candidates for interactive quantum field theory 
models are gauge theories. Scalar or multi-component scalar 
field theories of $P(\phi)_4$-type are less promising 
to describe interaction, since their construction either  
run into difficulties with renormalizability 
or, as conjectured for the $\phi_4^4$-model, they seem to be trivial
\cite{Froh82}. 

The description of gauge theories within the 
Wightman framework leads to some conceptional problems. 
For example, 
in order to study gauge invariant objects in quantum electro dynamics
one may think of 
vacuum expectation values of products of the field strength $F_{\mu\nu}$
\beqa
\6W_{\mu_1\nu_1,\cdots,\mu_n\nu_n}(x_1,\cdots, x_n)
=\<\Om,F_{\mu_1\nu_1}(x_1)\cdots F_{\mu_n\nu_n}(x_n)\Om\>
\eeqa
which satisfy the Wightman axioms. Here, the problem arises 
when one wish to include fermions. For the minimal coupling 
one has to study correlation functions of the gauge field 
instead of those in the field strength. This leads to such well 
known problems as indefinite metric, solving constraints and so forth.

Moreover, there is another problem which we would like to mention here.
Within the Wightman framework
the quantized version of the 
gauge field $u_\mu$ is an operator valued distribution. 
On the other hand, the classical 
concept of a gauge field leads to the notion of a connection in a 
vector or principal bundle over some manifold which suggests to 
consider as gauge invariant objects Wilson loop variables 
\beqa
w_\gam[u]=\tr[\8{Pexp}(\smallint_\gam A)]
\eeqa
and string-like objects  
\beqa
s_\gam[u,\psi]=\bar\psi(r(\gam))\8{Pexp}(\smallint_\gam u)\psi(s(\gam))
\eeqa
where $\psi$ is a smooth section in an appropriate vector bundle and 
$\gam$ is an oriented path which starts at $s(\gam)$ 
and ends at $r(\gam)$.

Unfortunately, to 
express $w_\gam(u)$ in terms of Wightman fields leads to difficulties.  
From a perturbation theoretical point of view one expects 
that the distribution $u$ is too singular in order to be 
restricted to a one-dimensional sub-manifold.  

To motivate our considerations, we shall discuss here, heuristically, 
an alternative proposal which might be related 
to a quantized version of a gauge theory.
It is concerned with the direct quantization of regularized  
Wilson loops 
\beqa
w_\gam(f)[u]=\int\8dx \ \ w_{\gam+x}[u] \ f(x) \ \ .
\eeqa
Here we allow $f\in E'(\7R^d)$ to be a 
distribution with compact support which has the form
\beqa
f(x)= f_\Sgm(x)\delta_\Sgm(x)
\eeqa
where $\Sgm$ is a $d-1$-dimensional hyper-plane and 
$f_\Sgm\in C^\infty_0(\Sgm)$ and $\delta_\Sgm$ is the natural measure 
on $\Sgm$. We claim that such a type 
of regularization is necessary since in    
$d$-dimensional quantum field theories there are no 
bounded operators which are localized within $d-2$-dimensional 
hyper-planes \cite{DrieSum86}. 

Such a point of view has  been 
discussed by J. Fr\"ohlich \cite{Froh79}, E. Seiler \cite{Seil82}
or more recently by A. Ashtekar and J. Lewandowski \cite{AshLew}.

In order to describe a quantum gauge theory in terms of regularized
Wilson loop variables one wishes to construct a
function $\gam\mapsto \1w_\gam$ which assigns to each 
path $\gam$ an operator valued distribution 
$\1w_\gam:f\mapsto \1w_\gam(f)$, where the operators $\1w_\gam(f)$
are represented by operators on some Hilbert space $\2H$. 
Heuristically, one expects that the operators $\1w_\gam(f)$
are unbounded \cite{Pol79}. 
\bdes
\itno 1
The operators $\1w_\gam(f)$  are self-adjoint for real-valued 
test functions with a joint core $\2D\subset \2H$
\itno 2
$\1w$ should transform 
covariantly under the action of the Poincar\' e group, i.e. 
\beqa
\1w_{g\gam}(f\circ g^{-1})= U(g)\1w_\gam(f) U(g)^* \ \ ; \ \ g\in\Poin\ \ ,
\eeqa
where $U$ is a unitary strongly continuous representation of 
the Poincar\' e group on $\2H$ and the spectrum of the 
translations is contained in the closed forward light cone $\bar V_+$.
\itno 2
Moreover, the operators  $\1w_\gam(f)$ should satisfy the locality 
requirement, i.e.  
\beqa
[\1E_{(\gam,f)}(\Delta_1), \1E_{(\gam_1,f_1)}(\Delta)]=0
\eeqa
if the (convex hulls) of the regions $\gam+\supp(f)$ 
and $\gam_1+\supp(f_1)$ are space-like separated. Here
\beqa
\1w_\gam(f)&=&\int\8d \1E_{(\gam,f)}(\lam) \ \lam 
\eeqa
\edes
is the spectral resolution of $\1w_\gam(f)$

According to \cite{Froh79,Seil82},  
it has been suggested to reconstruct Wilson loop operators 
$\1w_\gam$ from euclidean correlation functions of loops 
\beqa
\gam_1,\cdots, \gam_n\rMapsto\6S_n(\gam_1,\cdots, \gam_n)
\eeqa
which satisfy the analogous axioms as the usual 
Schwinger distributions do, namely the reflexion 
positivity and the symmetry. However, 
within the analysis of J. Fr\"ohlich, K. Osterwalder and 
E. Seiler \cite{FrohOstSeil,Seil82}, 
the correlation function may have singularities 
in those points where two loops intersect and 
there are some additional technical conditions assumed which are 
related to the behavior of these singularities. He has 
proven (compare also \cite{Froh79}) that one can reconstruct from the 
euclidean correlation functions $\6S_n$ 
an operator valued function $\gam\mapsto \1w_\gam$ 
together with a unitary strongly continuous representation of 
$\Poin$ on $\2H$ \cite{FrohOstSeil}. Here $\1w_\gam$ is only defined for 
loops which are contained in some space-like plane and it 
fulfills the covariance condition {\it (1)}. 
E. Seiler \cite{Seil82} has also discussed an idea how to 
proof locality {\it (2)}. We shall come back to this point later.

For our purpose, we look from an algebraic point of view 
at the problem of reconstructing 
a quantum field theory from euclidean data. Let us consider 
functions  
\beqa
a:\2A_E\ni u\rMapsto 
a^\circ \biggl( \int \8d x \ w_{\gam_j+x}(u) \ f_j(x); j=1,\cdots, n\biggr)
\eeqa
on the space of smooth connections $\2A_E$ in a vector bundle $E$ 
over the euclidean space $\7R^d$ where $a^\circ$ is a bounded function 
on $\7R^n$. These functions are bounded 
and thus they generate an abelian C*-algebra $A$ with C*-norm 
\beqa
\|a\|=\sup_{u\in\2A_E}|a(u)| \ \ .
\eeqa
We assign to a given 
bounded region $\2U\subset\7R^d$ the C*-sub-algebra $A(\2U)\subset A$ 
which is generated by all functions of Wilson loop variables $w_\gam(f)$ 
with $\gam+\supp(f)\subset\2U$. The euclidean group $\8E(d)$ acts 
naturally by automorphisms on $A$, namely the prescription 
\beqa
\al_g:a\rMapsto a\circ g^{-1}:u\rMapsto a(u\circ g) 
\eeqa
defines for each $g\in\8E(d)$ an appropriate automorphism of $A$, which, 
of course,  acts covariantly on the isotonous net 
\beqa
\ul A:\2U\rMapsto A(\2U) \ \ ,
\eeqa
namely we have: $\al_gA(\2U)=A(g\2U)$. 

Motivated by the work of E. Seiler, J. Fr\"ohlich 
and K. Osterwalder \cite{Seil82,Froh79,FrohOstSeil} 
as well as that of A. Ashtekar and J. Lewandowski \cite{AshLew}, 
we propose to 
consider reflexion positive functionals on $A$, i.e. linear functionals 
$\eta\in A^*$ which fulfill 
conditions, corresponding to the axioms {\it (E0)}-{\it (E2)} above.
These functionals can be interpreted as the analogue of the 
functional integral. 
Note, if $\eta$ is a state, then $\eta$ is nothing else but a measure 
on the spectrum $X$ of the C*-algebra $A$.  
The advantage of this point of view is based on the fact that 
abelian C*-algebras are rather simple objects namely 
algebras of continuous functions on a (locally)-compact 
Hausdorff space. 

\paragraph{Overview.}
In order to make the comprehension of the subsequent sections 
easier, we shall give an overview of the content of our paper
by stating the main ideas and results. This paragraph is also addressed 
to quick readers who are not so much interested into   
technical details. 

Motivated by the considerations above, we make 
in Section \ref{axioms} a suggestion for axioms which an 
euclidean field theory should satisfy. 
We start from an isotonous net 
\beqa
\ul A:\2U\rMapsto A(\2U)\subset A
\eeqa
of C*-algebra on which 
the euclidean group $\8E(d)$ acts covariantly by 
automorphisms of $\al:\8E(d)\to\aut A$, like in the example 
of Wilson loop variables given in the previous paragraph. 
However, we assume a somewhat weaker condition than commutativity 
for $A$. For our considerations we only have to assume 
that two operators commute if they are localized in 
disjoint regions. 
In addition to that, we consider a 
reflexion positive functional $\eta$ on $A$. 
We shall call the triple $(\ul A,\al,\eta)$, consisting of 
the net $\ul A$ of C*-algebra, the action of the 
euclidean group $\al$, and the reflexion positive functional, 
an euclidean field.  
   
We show in Section \ref{efthqfth} how to construct from 
a given euclidean field a quantum field theory in a 
particular vacuum representation. 
In order to point out the relation between the euclidean 
field $(\ul A,\al,\eta)$ and the Minkowskian world, 
we briefly describe the construction of a 
Hilbert space $\2H$ on which the reconstructed physical 
observables are represented.
According to our axioms, the map 
\beqa
a\otimes b\rMapsto <\eta,\iota_e(a^*)b>
\eeqa
is a positive semidefinite sesqui-linear form on the algebra $A(e)$
of operators which are localized in $e\7R_++\Sgm_e$ where 
$\Sgm_e$ is the hyper plane orthogonal to the 
euclidean time direction $e\in S^{d-1}$. Here $\iota_e$ is 
the automorphism on $A$ which corresponds to the reflexion
$e\mapsto -e$. By dividing the null-space and taking the closure 
we obtain a Hilbert space $\2H$. The construction 
of the observables, which turn out to be bounded operators on $\2H$, 
is based on two main steps.

\subparagraph{\it Step 1:} 
In Section \ref{reconpoin}, we reconstruct a 
unitary strongly continuous representation of the Poincar\' e group 
$U$ on $\2H$. To carry through this analysis, it is not 
necessary to impose new ideas. The construction is essentially analogous 
to those which has been presented in \cite{FrohOstSeil}
(compare also \cite{Seil82}). 
In order to keep the present paper self contained, 
we feel obliged to discuss this point within our context in more detail. 

\subparagraph{\it Step 2:} 
We discuss in Section \ref{reconlocobs} 
the construction of the physical observables. 
At the moment this can only be done, if we assume that 
the algebra $A$ contains operators which are localized 
at sharp times, i.e. we require that the algebra $A(e)\cap A(-e)$  
is larger than $\7C\11$. We shall abbreviate this condition by 
(TZ) which stands for {\em time-zero}. For the fix-point algebra $B(e)$ of 
$\iota_e$ in $A(e)\cap A(-e)$ we obtain a *-representation $\pi$ on 
$\2H$, where an operator $\pi(b)$, $b\in B(e)$, is given by 
the prescription 
\beqa
\pi(b)\8p(a)\rMapsto \8p(ba) \ \ .
\eeqa
Here $\8p$ is the 
quotient map, identifying an operator $a\in A(e)$ with its 
equivalence class $\8p(a)$ in $\2H$. Now, we consider 
for a given Poincar\' e transform $g\in\Poin$ and a given 
time-zero operator $b\in B(e)$ the following bounded 
operator:
\beqa
\Phi(g,b)&:=&U(g)\pi(b)U(g)^* \ \ .
\eeqa
We shall say that $\Phi(g,b)$ is localized in a 
region $\2O$ in Minkowski space if 
$b$ is localized in $\2U\subset \Sgm_e$ and 
the transformed region $g\2U$ is contained in
the double cone $\2O$.
Let us denote the C*-algebra which is 
generated by all operators $\Phi(g,b)$, which are localized in 
$\2O$, by $\6A(\2O)$. 
Hence we get an isotonous net of C*-algebras
\beqa
\ul{\6A}:\2O\rMapsto\6A(\2O)
\eeqa
indexed by double cones in Minkowski space on which 
the Poincar\' e group acts covariantly by the automorphisms 
$\al_g:=\Ad(U(g))$, $g\in\Poin$.    

\subparagraph{\it The main result:}
\bdes
\itno 1
The reconstructed isotonous net $\ul{\6A}$ is a
Haag-Kastler net: locality holds, 
i.e. if $\2O,\2O_1$ are two double cones such that 
$\2O\subset\2O_1'$ then $[\6A(\2O),\6A(\2O_1)]=\{0\}$.
\itno 2
Furthermore, 
the $\Poin$-invariant vector $\Om=\8p(\11)$ induces 
a vacuum state 
\beqa
\om:a\rMapsto<\om,a>&:=&\<\Om,a\Om\> \  \ .
\eeqa
\edes

The non trivial aspect of this statement is the proof of locality.
As already mentioned above, E. Seiler has discussed an 
idea how to prove locality   for a net of Wilson loops $\1w_\gam$.
This idea does not rely on the fact that one considers loops.
It can also be used for general euclidean fields. 
However, we have not found a complete proof within the common
literature and therefore, which is also one purpose of our paper,  
we shall present a complete proof
here (Section \ref{reconlocobs}). 
The prove is based on the analytic properties of the functions 
\beqa
F(z_1,z_2)&:=&\<\psi,\Phi_{X_1}(z_1,b_1)\Phi_{X_2}(z_2,b_2)\hat\psi\> 
\vs\vs
\hat F(z_1,z_2)&:=&\<\psi,\Phi_{X_2}(z_2,b_2)\Phi_{X_1}(z_1,b_1)\hat\psi\>
\ \ .
\eeqa
We have introduced the operators 
\beqa
\Phi_X(z,b):=U(\exp(zX))\pi(b)U(\exp(-zX))
\eeqa
where $b\in B(e)$ is a time-zero operator and $X$ is a Boost generator 
or $\8iH$ where $H$ is the hamiltonian with respect to the 
time direction $e$.

Roughly, the argument for the proof of locality goes as follows:
Suppose $b_j$ is localized in $\2U_j\subset\Sgm_e$. 
We shall show that the regions $G$ ($\hat G$) in which $F$ ($\hat F$) 
are holomorphic are 
\bdes
\itno a connected and they contain pure imaginary points $(\8is_1,\8is_2)$ and
\itno b the intersection $G\cap\bar G$ contains all those points $(t_1,t_2)$ 
for which $\2O_1=\exp(t_1X)\2U_1$ and $\2O_2=\exp(t_2X_2)\2U_2$ are 
space-like separated.
\edes
But $F$ and $\hat F$ coincide in the pure imaginary points
since operators which are localized in disjoint regions commute.
This implies 
\beqa
F|_{G\cap\hat G}=\hat F|_{G\cap\hat G}
\eeqa
and thus by {\it (b)} we conclude 
\beqa
\<\psi,[\Phi_{X_1}(t_1,b_1),\Phi_{X_2}(t_2,b_2)]\hat\psi\>&=&0
\eeqa
if $\Phi_{X_1}(t_1,b_1)$ and  $\Phi_{X_2}(t_2,b_2)$ are localized in 
space-like separated regions. We claim that the regions
$G$ and $\hat G$ depend on the choice of the vector
$\hat\psi$. However, one can find a dense sub-space $D$
such that $F$ ($\hat F$) are holomorphic in 
$G$ ($\hat G$) for all $\hat\psi\in D$. Thus the 
commutator $[\Phi_{X_1}(t_1,b_1),\Phi_{X_2}(t_2,b_2)]$ vanishes 
on a dense sub-space and, since $\Phi_X(t,b)$ is bounded 
for real points $t\in\7R$, the commutator vanishes on $\2H$. 

In order to get analyticity of $F$ within a region $G$ which is large enough, 
we prove in the appendix an statement which is the analogue 
of the famous Bargmann-Hall-Wightman theorem 
\cite{HallWght57,Jost65,StrWgh89}.  

In Section \ref{strucasp}, we discuss some miscellaneous consequences 
of our result. Note, that for the application of our reconstruction
scheme it was crucial to assume that the there are 
non-trivial euclidean operators which can be localized at sharp times. 
We shall give some remarks on the condition (TZ) in  
Section \ref{tz}. 

Our considerations can easily be generalized to 
the case in which there are also fermionic operators present or even though 
for super-symmetric theories. 
Here one starts with an isotonous net 
$\ul F:\2U\mapsto F(\2U)$ of $\7Z_2$-graded C*-algebras
which fulfills the time-zero condition (TZ), i.e. the 
fix-point algebra $B(e)$ of $\iota_e$ in $F(e)\cap F(-e)$   
is larger than $\7C\11$. The euclidean group acts covariantly by automorphisms 
on $F$ and we require that the graded commutator 
$[a,b]_g=0$ vanishes if $a$ and $b$ are localized in 
disjoint regions. 

Let $\eta$ be a reflexion positive functional, then, by 
replacing the commutator by the graded commutator, 
we conclude that the operators 
\beqa
\Phi(g,b) = U(g)\pi(b)U(g)^* \ \ \mbox{ ; $b\in B(e)$ and $g\in\Poin$} 
\eeqa
generate a fermionic net $\ul{\6F}$ of C*-algebras.
This can really be done analogously to the construction of 
the Haag-Kastler net $\ul{\6A}$, described above. 

Finally, we close our paper by the Section \ref{conout} 
{\em conclusion and outlook}.

\section{Axioms for euclidean field theories}
\label{axioms}
In the present section we make a suggestion for axioms which an euclidean 
field theory should satisfy. 

In the first step, we 
introduce the notion of an {\em euclidean net of C*-algebras}. 
Within our interpretation 
this notion is related to {\em physical observations}.

\bdef
A $d$-dimensional {\em euclidean net} of C*-algebras is given by a 
pair $(\ul A,\al)$ which consists of an isotonous net 
\beqa
\ul A:\7R^d\supset\2U\rMapsto A(\2U)
\eeqa
of C*-algebras, indexed by bounded subsets  
in $\7R^d$ and   
a group homomorphism $\al\in\hom(\8E(d),\aut(A))$.\footnote{We denote the  
the C*-inductive limit of $\ul A$ by $A$. For an unbounded region $\Sgm$ 
the algebra $A(\Sgm)$ denotes the C*-sub-algebra which is generated by 
the algebras $A(\2U)$, $\2U\subset \Sgm$.} 
We require that the pair fulfills the conditions: 
\bdes
\itno 1 Locality: $\2U_1\cap\2U_2=\emptyset$ implies 
$[A(\2U_1),A(\2U_2)]=\{0\}$. 
\itno 2 Euclidean covariance: $\al_gA(\2U)=A(g\2U)$ for each $\2U$.
\edes
\eef

For an euclidean direction 
$e\in S^{d-1}$ we consider the reflection $\te_e:e\mapsto -e$.
and the sub-group $\8E_e(d-1)$ which commutes with $\te_e$.
Moreover, we set $\iota_e:=\al_{\te_e}$. 
As in the introduction, we denote by $A(e)$ the C*-algebra 
$A(e\7R_++\Sgm_e)$ where $\Sgm_e$ is the hyper-plane orthogonal to $e$.   

Now we formulate a selection criterion for linear functionals 
on $A$ which corresponds to the selection criterion for 
physical states. We shall see that class of functional, which is introduced 
below, is the euclidean analogue of the set of vacuum states.

\bdef
We define $S(A,\al)$ to be the set of all continuous linear functionals $\eta$ 
on $A$ which fulfill the following conditions:
\bdes
\itno 1 
$e$-reflexion positivity: There exists a euclidean direction $e\in S^{d-1}$ 
such that   
\beqa
\forall a\in A(e): &<\eta,\iota_e(a^*)a>& \geq  0  \ \ . 
\eeqa
\itno 2
Unit preserving: $<\eta,\11>=1$.
\itno 3
Invariance: $\forall g\in \8E(d): \eta\circ\al_g=\eta$. 
\edes
\eef

\paragraph{Remark:}
We easily observe that the definition of  
$S(A,\al)$ is independent of the chosen direction $e$. 
In the subsequent, we call the functionals in $S(A,\al)$ 
{\em reflexion positive}.
\skb

For our purpose it is necessary to require a further condition for 
the functionals under consideration.

\bdef
We denote by $S_R(A,\al)$ the set of all 
reflexion positive functionals $\eta$ of $A$ for which the  
map  
\beqa
\8E(d)\ni g\rMapsto <\eta,a(\al_gb)c>
\eeqa
is a continuous function for each $a,b,c\in A$.
These functionals are called {\em regular reflexion positive}.
\eef

We shall call a triple $(\ul A,\al,\eta)$ which consists of 
an euclidean net and a regular reflexion positive 
functional $\eta$ an {\em euclidean field}.

As already mentioned in the introduction, we have to assume that the 
operators of the euclidean net can be localized at a sharp 
$d-1$-dimensional hyper plane.  For an euclidean time direction $e$ 
we denote by $B(e)$ the fix-point algebra of 
$A(e)\cap A(-e)$ under the reflexion $\iota_e$. 

\paragraph{Condition (TZ):} 
A $d$-dimensional euclidean net of C*-algebras $(\ul A,\al)$ 
fulfills the time-zero condition (TZ) iff $B(e)$ is a non-trivial 
C*-algebra, i.e. it is not $\7C\11$. 
We call the algebras $B(e)$ time-zero algebras. 
For a region $\2U\subset \Sgm_e$, we denote by $B(e,\2U)$ the 
sub-algebra which is generated by operators localized 
in $\2U$.
\skb

\paragraph{Remark:}
Let $(\ul A,\al)$ be a $d$-dimensional euclidean net of C*-algebras
which fulfill the condition (TZ). Then 
the net 
\beqa
\ul B^e:\Sgm_e\supset \2U\rMapsto B(e,\2U)
\eeqa
together with the group homomorphism 
$\be^e:=\al|_{E_e(d-1)}$ is, of course, 
a $d-1$-dimensional euclidean net of C*-algebras.

\section{From euclidean field theory to quantum field theory}
\label{efthqfth}

In the present section, we discuss how to pass from a 
euclidean field $(\ul A,\al,\eta)$ to a quantum 
field theory in a particular vacuum representation. 

In the first step we construct from a given euclidean field
$(\ul A,\al,\eta)$ a unitary strongly continuous 
representation of the Poincar\' e group (Section \ref{reconpoin}).

In the second step we have to require that  condition (TZ) is satisfied
in order to show that a concrete Haag-Kastler can be reconstructed from 
the elements of the time-zero algebras and the representation of 
the Poincar\' e group (Section \ref{reconlocobs}).

\subsection{Reconstruction of the Poincar\'e group}
\label{reconpoin}

For $e\in S^{d-1}$ we introduce a positive semidefinite sesqui-linear form on
$A(e)$ as follows:
\beqa
a\otimes b \rMapsto <\eta,\iota_e(a^*)b> \ \ .
\eeqa
Its null space is given by 
\beqa
N(e,\eta):=\{a\in A(e)|\forall b\in A(e):<\eta,\iota_e(a^*)b>=0 \} 
\eeqa
and we obtain a pre-Hilbert space 
\beqa
D(e,\eta):=A(e)/N(e,\eta)
\eeqa
The corresponding quotient map is denoted by 
\beqa
\8p_{(e,\eta)}:A(e)\rTo D(e,\eta) 
\eeqa
and its closure $\2H(e,\eta)$ is a Hilbert space with scalar product
\beqa
\<\8p_{(e,\eta)}(a),\8p_{(e,\eta)}(b)\>&:=&<\eta,\iota_e(a^*)b> \ \ .
\eeqa

\blem\label{lem1}
The map 
\beqa
T_{(e,\eta)}:s\in\7R_+ \rMapsto T_{(e,\eta)}(s):
\8p_{(e,\eta)}(a) {\rMapsto} \8p_{(e,\eta)}(\al_{(1,se)}a) 
\eeqa
is a strongly continuous semi-group of contractions with a positive 
generator $H_{(e,\eta)}\geq 0$. 
\elem
\bpr
Since 
\beqa
<\eta,\iota_e(b^*)a> \ \ = \ \ 0
\eeqa
for each $b\in A(e)$ implies
\beqa
<\eta,\iota_e(b^*)\al_{se}a> \ \ = \ \ <\eta,\iota_e(\al_{se}b^*)a>=0
\eeqa
for each 
$b\in A(e)$, we conclude that 
\beqa
T_{(e,\eta)}(s)\8p_{(e,\eta)}(a)=0
\eeqa
for $a\in N(e,\eta)$. Hence $T_{(e,\eta)}$ 
is well defined. 
The fact that $T_{(e,\eta)}$ is a semi-group of contractions 
follows by standard arguments, 
i.e. a multiple application of the Cauchy-Schwartz inequality. 
Finally, the strong continuity follows from the regularity 
of $\eta$.
\epr

We consider the set $\8{Con}(e)$ of all cones  $\Gam$ (in euclidean space)
of the form $\Gam=\7R_+(B_d(r+e))+\eps e$ 
where $B_d(r)$ denotes the ball in $\7R^d$ with center $x=0$ and radius $r$.
In addition, we define the following subspace of $\2H(e,\eta)$
\beqa
D(\Gam;\eta):=\8p_{(e,\eta)}A(\Gam) \ \ .
\eeqa

\blem\label{lem2}
For each cone $\Gam\in\8{Con}(e)$, 
the vector space $D(\Gam,\eta)$ is a dense subspace of $\2H(e,\eta)$.
\elem
\bpr
Lemma \ref{lem1} states that 
$T_{(e,\eta)}$
is a semi-group of contractions with a positive generator. Furthermore, 
$D(\Gam,\eta)$ is mapped into itself by $T_{(e,\eta)}(s)$. 
Since for each operator $a\in A(e)$ there exists an $s>0$
such that 
\beqa 
T_{(e,\eta)}(s)\8p_{(e,\eta)}(a)\in D(\Gam,\eta) \ \ , 
\eeqa
we can apply a  Reeh-Schlieder argument in order to prove that $D(\Gam,\eta)$
is a dense subspace of $\2H(e,\eta)$.
\epr

\blem\label{lem3}
Let $\2V\subset\8E(d)$ be a small neighborhood of 
the unit element $1\in\8E(d)$ and let $\Gam\in\8{Con}(e)$ be a cone 
such that $\2V\Gam\subset e\7R_++\Sgm_e$.
Then $a\in A(\Gam)\cap N(e,\eta)$ implies $\al_ga\in N(e,\eta)$ 
for each $g\in \2V$.
\elem
\bpr
We have $<\eta,\iota_e(b^*)\al_{se}a> \ \ = \ \ 0$
for each $b\in A(\Gam)$ and hence 
$<\eta, \iota_e(b^*)\al_ga>=<\eta,\iota_e(\al_{\te_eg}b^*)a>=0$.
Since we may choose $\2V$ to be $\te_e$-invariant, we have 
$\al_{\te_eg}b^*\in A(e)$ and the result follows by 
Lemma \ref{lem2}.
\epr

\bthe\label{thepoin} 
Let $\eta\in S_R(A,\al)$ be a regular reflexion positive functional. 
Then for each 
$e\in S^{d-1}$ there exists a unitary strongly continuous 
representation $U_{(e,\eta)}$
of the $d$-dimensional Poincar\'e group $\Poin$ 
\beqa
U_{(e,\eta)}\in\hom[\Poin, U(\2H(e,\eta))] 
\eeqa
such that the spectrum of the translations $x\to U_{(e,\eta)}(1,x)$ 
is contained in the closed forward light cone $\bar V_+$.
\ethe

\bpr
The theorem 
can be proven by using the proof of \cite[Theorem 8.10]{Seil82}. 
We briefly illustrate the construction of the representation 
$U_{(e,\eta)}$. Let $\2V\subset\8E(d)$ be a small neighborhood of 
the unit element $1\in\8E(d)$. Then there exists a 
cone $\Gam\in\8{Con}(e)$  such that $\2V\Gam\subset e\7R_++\Sgm_e$. 
According to Lemma \ref{lem3} we may define for each 
$g\in \2V$ the operator
\beqa
V_{(e,\eta)}(g)\8p_{(e,\eta)}(a):=\8p_{(e,\eta)}(\al_ga)
\eeqa
with domain $D(\Gam,\eta)$.
If $g$ belongs to the group $\8E_e(d-1)$ then 
we conclude that $V_{(e,\eta)}(g)=U_{(e,\eta)}(g)$ is a unitary operator.

Let $\6e(d)$ be the Lie algebra of $\8E(d)$
and let $\6e_e(d-1)\subset\6e(d)$ be the 
sub-Lie algebra of $\8E_e(d-1)\subset \8E(d)$. We decompose 
$\6e(d)$ as follows:
\beqa
\6e(d)=\6e_e(d-1)\oplus\6m_e(d-1)
\eeqa
and we obtain another real Lie algebra:
\beqa
\6p(d):=\6e_e(d-1)\oplus\8i\6m_e(d-1)
\eeqa
which is the Lie algebra of the Poincar\'e group $\Poin$.

For each $X\in\6m_e(d-1)$ there 
exists a self adjoint operator $L_{(e,\eta)}(X)$ where $D(\Gam,\eta)$
consists of analytic vectors for $L_{(e,\eta)}(X)$ and 
for each $s\in\7R$ with $\exp(sX)\in \2V$ we have:
\beqa
V_{(e,\eta)}(\exp(sX))=\exp(sL_{(e,\eta)}(X)) \ \ .
\eeqa

According to \cite[Theorem 8.10]{Seil82} we conclude that 
the unitary operators 
\beqa
U_{(e,\eta)}(\exp(\8isX)):=\exp(\8isL_{(e,\eta)}(X)) &;& X\in\6m_e(d-1)  
\vs\vs
U_{(e,\eta)}(g):=V_{(e,\eta)}(g) &;& g\in\8E_e(d-1)
\eeqa
induce a unitary strongly continuous representation of the 
Poincar\'e group $\Poin$. The positivity of 
the Energy follows from the positivity of the transfer matrix
$T_{(e,\eta)}(1)$.  
\epr

\paragraph{Remark:}
The vector $\Om_{(e,\eta)}:=\8p_{(e,\eta)}(\11)$ is invariant under 
the action of the Poincar\'e group.

\subsection{Reconstruction of the net of local observables}
\label{reconlocobs}

In the subsequent, 
we consider a euclidean net of C*-algebras $(\ul A,\al)$ which 
fulfills the condition (TZ).

\bpro
Let $\eta$ be a regular reflexion positive functional
on $A$. Then the map 
\beqa
\pi_{(e,\eta)}:B(e)\ni b \rMapsto \pi_{(e,\eta)}(b):
\8p_{(e,\eta)}(a) \rMapsto \8p_{(e,\eta)}(ba)
\eeqa
is a well defined *-representation of $B(e)$. 
\epro
\bpr
For each $a\in N(e,\eta)$ and for each $c\in A(e)$ we have
\beqa
<\eta,\iota_e(c^*)ba> \ \ = \ \ <\eta,\iota_e(c^*b)a> \ \ = \ \ 0
\eeqa
and hence $\pi_{(e,\eta)}(b)$ is a well defined linear and bounded operator. 
By construction it is clear 
that $\pi_{(e,\eta)}$ is a *-homomorphism. 
\epr

\paragraph{Remark:}
The restriction of $\eta|_{B(e)}$ 
is a state of $B(e)$.  Of course, the GNS-representation of 
$\eta|_{B(e)}$ is a sub-representation of $\pi_{(e,\eta)}$.

\bdef
\bdes
\itno 1
Let $\2O$ be a double cone in $\7R^d$. Then we define  
$\6A_{(e,\eta)}(\2O)$ to be the C*-algebra on 
$\2H(e,\eta)$ which is generated by operators 
\beqa
\Phi_{(e,\eta)}(g,b):=U_{(e,\eta)}(g)\pi_{(e,\eta)}(b)U_{(e,\eta)}(g)^* 
\eeqa
with $b\in B(e,\2U)$, $g\in\Poin$ and $g\2U\subset\2O$.
\itno 2
We denote by $\ul{\6A}_{(e,\eta)}$ the net of C*-algebras which is 
given by the prescription
\beqa
\ul{\6A}_{(e,\eta)}:\2O\rMapsto \6A_{(e,\eta)}(\2O) \ \ .
\eeqa
\edes
\eef

\bthe\label{thehk}
The pair $(\ul{\6A}_{(e,\eta)},\Ad(U_{(e,\eta)}))$
is a $\Poin$-covariant Haag-Kastler which is represented 
on $\2H(e,\eta)$. 
\ethe

\paragraph{Remark:} 
\bdes
\itno 1
Note that 
\beqa
\om_{(e,\eta)}:\6A_{(e,\eta)}\ni a\rMapsto \<\Om_{(e,\eta)},a\Om_{(e,\eta)}\>
\eeqa
is a vacuum state since $U_{(e,\eta)}$ is a positive energy 
representation of the  Poincar\'e group.
However, in general $\om_{(e,\eta)}$ is not a pure state.
\itno 2
For the local algebra $\6A_{(e,\eta)}(\2O)$, we do not 
take the von Neumann algebra generated by the corresponding 
operators $\Phi_{(e,\eta)}(g,b)$ since this might to problems 
with locality.
\edes

\paragraph{Preparation of the proof of Theorem \ref{thehk}:}
For a Lie algebra element $X\in\8i\6m_e(d-1)$ and a complex 
number $z\in\7C$ we define a linear (unbounded) operator on $\2H(e,\eta)$ by
\beqa
\Phi_{(e,\eta,X)}(z,b):=
U_{(e,\eta)}(\exp(zX))\pi_{(e,\eta)}(b)U_{(e,\eta)}(\exp(-zX)) 
\eeqa
on a dense domain $D(\Gam,\eta)$ where $\Gam\in\8{Con}(e)$
an appropriate cone. 

In order to formulate the our next result, we define for two generators 
$X_1,X_2\in\8i\6m_e(d-1)$, for an interval $I$, for a
neighborhood $\2V\supset\Lor$ of the unit element in $\8P_+(\7C)$, and for 
two subsets $\2U_j\subset\Sgm_e$, $j=1,2$, the region
\beqa
G(\2V;X_1,X_2;\2U_1,\2U_2;I)&:=&\bigcup_{g\in\2V\times\Lor}
\biggl\{(z_1,z_2)\in(\7R\times\8iI)^2\biggm |\forall \1x_j\in\2U_j:
\vs\vs
&&e \ \8{Im}[g(\exp(z_1X_1)\1x_1-\exp(z_2X_2)\1x_2)]\in \7R_+ \biggr\} \ \ .
\eeqa
We shall prove in the appendix the lemma given below
which is the analogue of the famous BHW theorem (compare also 
\cite{Jost65,StrWgh89}
and references given there):

\blem\label{lem21}
For a given interval $I$, 
there exists a dense subspace $D\subset\2H(e,\eta)$, such that the function 
\beqa
F_{(X_1,X_2,b_1,b_2)}:(z_1,z_2)\rMapsto 
\<\psi_1,\Phi_{(e,\eta,X_1)}(z_1,b_1)\Phi_{(e,\eta,X_2)}(z_2,b_2)\psi_2\>
\eeqa
is holomorphic in $G(\2V;X_1,X_2;\2U_1,\2U_2,I)$ 
for each $\psi_1,\psi_2\in D$.
\elem
 
We claim that the $E(d)$ invariance of $\eta$ yields that 
the dense subspace $D\subset\2H(e,\eta)$ can be chosen in such a way that
\beqa
&&\8I(\2V;X_2,X_2;\2U_2,\2U_1;I)
\vs\vs
&&:= G(\2V;X_2,X_2;\2U_2,\2U_1;I)\cap 
G(\2V;X_1,X_2;\2U_1,\2U_2;I)\cap\8i\7R^2
\vs\vs
&&\not= \emptyset \ \ .
\eeqa

\blem\label{lem22}
If $\2U_1\cap\2U_2=\emptyset$ and 
$(s_1,s_2)\in \8I(\2V;X_2,X_2;\2U_2,\2U_1;I)$, then 
\beqa
F_{(X_1,X_2,b_1,b_2)}(\8i s_1,\8i s_2)=
F_{(X_2,X_1,b_2,b_1)}(\8i s_2,\8i s_1) \ \ .
\eeqa
\elem
\bpr
The lemma is a direct consequence of the euclidean covariance and the 
locality of the net $\ul A$.
\epr

\paragraph{\it Proof of Theorem \ref{thehk}:} 
We conclude from Theorem \ref{thepoin} and the construction of the 
algebras $\6A_{(e,\eta)}(\2O)$ that $\ul{\6A}_{(e,\eta)}$
is a Poincar\'e covariant net of C*-algebras, represented on 
$\2H(e,\eta)$.  

It remains to be proven that $\ul{\6A}_{(e,\eta)}$ is a local net.
For this purpose it is sufficient to show that for each pair 
\beqa
(t_1,t_2)&\in& R(X_1,X_2;\2U_1,\2U_2) 
\vs\vs
&:=&\{ (t_1,t_2) \in\7R^2 |\exp(t_1X_1)\2U_1\subset(\exp(t_2X_2)\2U_2)'\} 
\eeqa
the commutator  
\beqa
[\Phi_{(e,\eta,X_1)}(t_1),\Phi_{(e,\eta,X_2)}(t_2)]|_D=0
\eeqa
vanishes on an appropriate dense domain $D\subset\2H(e,\eta)$.

Since the points in $R(X_1,X_2;\2U_1,\2U_2)$ are space-like points, 
we conclude that there exist complex 
Lorenz boosts $g_\pm\in\2V$ such that 
\beqa
\8{Im}g_\pm R(X_1,X_2;\2U_1,\2U_2))\subset V_\pm \ \ .
\eeqa
Hence we have 
\beqa
R(X_1,X_2;\2U_1,\2U_2)\subset G(\2V;X_1,X_2;\2U_1,\2U_2;I)\cap
G(\2V,X_2,X_1;\2U_2,\2U_1;I) \ \ .
\eeqa
Using Lemma \ref{lem22}, we conclude that 
\beqa
F_{(X_1,X_2,b_1,b_2)}(z_1,z_2)=F_{(X_2,X_1,b_2,b_1)}(z_2,z_1)
\eeqa
for 
\beqa
(z_1,z_2)\in G(\2V;X_1,X_2;\2U_1,\2U_2;I)\cap
G(\2V;X_2,X_1;\2U_2,\2U_1;I)
\eeqa
which finally yields 
\beqa
F_{(X_1,X_2,b_1,b_2)}(t_1,t_2)=F_{(X_2,X_1,b_2,b_1)}(t_2,t_1)
\eeqa
for each $(t_1,t_2)\in R(X_1,X_2;\2U_1,\2U_2)$. This proves the locality 
of $\ul{\6A}_{(e,\eta)}$.
\epr

\section{Discussion of miscellaneous consequences}
\label{strucasp}

Due to Theorem \ref{thehk} we are able to pass form a 
euclidean field $(\ul A,\al,\eta)$ to a quantum field theory 
in a particular vacuum representation. One crucial condition to apply
our method is the existence of the time-zero algebras.
We shall see that the discussion 
of Section \ref{tz} covers all possible situations 
for euclidean fields which fulfill the condition (TZ).

Afterwards, 
we discuss in Section \ref{fermi} how the 
reconstruction scheme has to be generalized in order to 
include fermionic operators.

\subsection{Some remarks on euclidean fields which satisfy the 
time-zero condition}
\label{tz}
Let us consider a $d-1$-dimensional euclidean net $(\ul B,\beta)$ of 
abelian C*-algebras.  
 
\bdef\label{defrel}
Let $G$ be a group which contains $\8E(d-1)$ as a sub-group. 
We define $A_0(G;B,\beta)$ to be the *-algebra 
which is generated by pairs $(g,b)\in G\times B$ modulo the relations:
\bdes
\itno 1
For each $g\in G$, the map $b \rMapsto (g,b)$ is a *-homomorphism.
\itno 2
For each $g\in G$, for each $h\in\8E(d-1)$, and for each $b\in B$:  
\newline $(gh,b)=(g,\beta_hb)$
\edes
\eef

The algebra $A_0(G;B,\beta)$ possesses a natural C*-norm which is given by 
\beqa
\|a\|:=\sup_{(\2H,\pi)\in R(G;B,\beta)}\|\pi(a)\|_{\2B(\2H)}
\eeqa
where $R(G;B,\beta)$ is the set of all representations $\pi$ of 
$A_0(G;B,\beta)$ by bounded operators on a Hilbert space $\2H$. 
The closure of $A_0(G;B,\beta)$ is denoted by $A(G;B,\beta)$. 

\paragraph{Remark:}
There is a natural group homomorphism \newline
$\al\in\hom(G,\aut A(G;B,\beta))$
and a natural faithful embedding $\phi\in\hom^*(B,A(G;B,\beta))$ given by:
\beqa
\al_g(g_1,b)&:=&(gg_1,b)
\vs\vs
\phi(b)&:=&(1,b) \ \ .
\eeqa
Of course, we have for each $h\in\8E(d-1)$:
\beqa
\phi\circ\beta_h&=&\al_h\circ\phi \ \ .
\eeqa
\skb

We are mostly interested in two cases for $G$, namely $G=\Poin$ and 
$G=\8E(d)$. For both groups $A(G;B,\beta)$ has a natural local structure
since $\Poin$ and $\8E(d)$ act as groups on $\7R^d$. 

\bdef
For a region $\2O\in\7R^d$ we define $A(G;B,\beta|\2O)$ to be the 
C*-sub-algebra in $A(G;B,\beta)$ which is generated by elements
$(g,b)$ with $b\in B(\2U)$ and $g\2U\subset \2O$ and we obtain nets 
\beqa
\ul A(G;B,\beta):\2O\rMapsto A(G;B,\beta|\2O) \ \ .
\eeqa
\eef

In order to get a Haag-Kastler net for $G=\Poin$ and a 
euclidean net for $G=\8E(d)$, we consider the following ideals:
\bdes
\itno 1
$J_c(\Poin;B,\beta)$ is the two-sided ideal which is generated by 
elements $[(g,b),(g_1,b_1)]$ where $(g,b)$ and $(g_1,b_1)$
are localized in space like separated regions.
\itno 2
$J_c(\8E(d);B,\beta)$ is the two-sided ideal which is generated by 
elements $[(g,b),(g_1,b_1)]$ where $(g,b)$ and $(g_1,b_1)$
are localized in disjoint regions.
\edes
Thus the prescription  
\beqa
\ul{\6A}_G:\2O\rMapsto \6A_G(\2O):=
A(G;B,\beta|\2O)/J_c(G;B,\beta)
\eeqa
is a $\Poin$-covariant Haag-Kastler net for $G=\Poin$, and
an euclidean net of C*-algebras for $G=\8E(d)$.

\bpro\label{prouni}
Let $(\ul A,\al)$ be a $d$-dimensional euclidean net which fulfills 
the condition (TZ) and let $(B,\be)$ be the $d-1$-dimensional 
euclidean net, corresponding to the hyper plane $\Sgm_e$. Then the map 
\beqa
\chi:\6A_{\8E(d)}\ni (g,b)\rTo \al_g(b)\in A
\eeqa
is a *-homomorphism which preserves indeed the net structure. 
\epro
\bpr 
By using the relations in 
Definition \ref{defrel} we conclude, by some 
straight forward computations, that $\chi$ is a 
a *-homomorphism which preserves the net structure.  
\epr 

An application of Theorem \ref{thehk} gives:

\bcor
For each regular reflexion positive functional $\eta$ on $\6A_{\8E(d)}$
there exists a vacuum state $\om_\eta$ on $\6A_\Poin$ such that 
\beqa
\om_\eta|_B=\eta|_B \ \ .
\eeqa
\ecor

\paragraph{Remark:}
\bdes
\itno 1
Note that we may view $B$ as a common sub algebra of $\6A_{\8E(d)}$ and 
$\6A_\Poin$ since $B\cap J_c(G;B,\beta)=\{0\}$.
\itno 2
Given an euclidean field $(\ul A,\al,\eta)$, for which the 
time zero algebra $B:=B(e)$ is non trivial. 
By Proposition \ref{prouni}, we conclude that there is a 
positive energy representation $\pi_{(e,\eta)}$ of 
$\ul{\6A}_\Poin$ on the Hilbert space $\2H(e,\eta)$ whose  
image is precisely the net $\ul{\6A}_{(e,\eta)}$. In particular 
the GNS-representation of $\om_\eta$ is a sub-representation of 
$\pi_{(e,\eta)}$.
\itno 3
Both, the algebra $\6A_\Poin$ of observables in 
Minkowski space and the euclidean algebra $\6A_{\8E(d)}$ 
can be considered as 
sub-algebras of $\6A_{P_+(\7C)}$ where the algebra  $\6A_{P_+(\7C)}$
is defined by 
\beqa
\6A_{P_+(\7C)}:=
A(P_+(\7C);B,\beta)/[J_c(\Poin;B,\beta)\cup J_c(\8E(d);B,\beta)] \ \ .
\eeqa
\edes

\subsection{The treatment of fermionic operators}
\label{fermi}

In order to discuss the treatment of fermionic operators we 
introduce the notion of a fermionic euclidean net. The 
axioms for such a net coincide with those of an
euclidean net, except the locality requirement. 

\bdef
An isotonous and $\8E(d)$-covariant net $(\ul F,\al)$  
\beqa
\ul F:\7R^d\supset\2U\rMapsto F(\2U)=F_+(\2U)\oplus F_-(\2U) 
\eeqa
of $\7Z_2$-graded C*-algebras is called a {\em fermionic euclidean net}
iff $\2U_1\cap\2U_2=\emptyset$ implies 
$[F(\2U_1),F(\2U_2)]_g=\{0\}$, where 
$[\cdot,\cdot]_g$ denotes the graded commutator. 
\eef

For a given $d-1$-dimensional fermionic net $(\ul F,\be)$, we 
build the C*-algebras $A(\8E(d);F,\beta)$ and $A(\Poin;F,\beta)$
as introduced in the previous section.
Note, that the algebra $A(\Poin;F,\beta)$ possesses a 
$\7Z_2$-grading, namely we have  
\beqa
A(\Poin;F,\beta)=A_+(\Poin;F,\beta)\oplus A_-(\Poin;F,\beta)
\eeqa
where the algebra $A_+(\Poin;F,\beta)$ is spanned  
by products of elements 
$(g,b)$ containing an even number of generators in $G\times F_-$:
\beqa
(g_1,b_1)\cdots (g_{2n},b_{2n}) \ \ .
\eeqa
Therefore the sub-space $A_-(\Poin;F,\beta)$
is spanned  by elements which are products of elements 
$(g,b)$ containing an odd number of generators in $G\times F_-$:
\beqa
(g_1,b_1)\cdots (g_{2n-1},b_{2n-1}) \ \ .
\eeqa

Analogously to the purely bosonic case, we consider the two-sided ideals
\bdes
\itno 1
$J_g(\Poin;F,\beta)$ which is generated by 
graded commutators $[(g,b),(g_1,b_1)]_g$ where $(g,b)$ and $(g_1,b_1)$
are localized in space like separated regions and 
\itno 2
$J_g(\8E(d);B,\beta)$ which is generated by 
graded commutators $[(g,b),(g_1,b_1)]_g$ where $(g,b)$ and $(g_1,b_1)$
are localized in disjoint regions.
\edes
Thus the prescription 
\beqa
\ul{\6F}_G:\2O\rMapsto\6F_G(\2O):= A(G;F,\beta|\2O)/J_g(G;F,\beta)
\eeqa
is a fermionic $\Poin$-covariant Haag-Kastler net for $G=\Poin$, and
an fermionic euclidean net for $G=\8E(d)$.

By following the arguments in the proof of Theorem \ref{thehk} 
and by keeping in mind that the ordinary commutator has to be 
substituted by the graded commutator, we get the result:

\bcor
For each regular reflexion positive functional $\eta$ on 
the fermionic euclidean net $\6F_{\8E(d)}$
there exists a vacuum state $\om_\eta$ on $\6F_\Poin$ such that 
\beqa
\om_\eta|_F=\eta|_F \ \ .
\eeqa
\ecor

\paragraph{Remark:} 
As described in 
Section \ref{reconlocobs} the state is defined by 
\beqa
<\om_\eta,\prod_{j=1}^n(g_j,b_j)>&=&
\<\Om_{(e,\eta)},\prod_{j=1}^n\Phi_{(e,\eta)}(g_j,b_j)\Om_{(e,\eta)}\> \ \ .
\eeqa


\section{Conclusion and outlook}
\label{conout}

\subsection{Concluding remarks and comparison}

We have shown, how a quantum field theory can be reconstructed 
form a given euclidean field $(\ul A,\al,\eta)$ which fulfills the 
condition (TZ). We think, that in comparison to the 
usual Osterwalder-Schrader reconstruction theorem the 
reconstruction of a quantum field theory from 
euclidean fields (in our sense) has the following advantages:

\paragraph{$\oplus$}
The Osterwalder-Schrader reconstruction theorem relates Schwinger 
distributions to a Wightman theory. 
One obtains an operator valued distribution 
$\Phi$ which satisfies the Wightman axioms. 
The reconstructed field operators $\Phi(f)$ are, 
in general, unbounded operators and in order to 
get a Haag-Kastler net of bounded operators one 
has to prove that not only the field operators 
$\Phi(f)$, $\Phi(f_1)$ commute if $f$ and $f_1$ have 
space-like separated supports, but also 
its corresponding spectral projections. Furthermore, as mentioned 
in the introduction, in order to apply the results of 
\cite{OstSchra1} one has to prove that the Schwinger distributions 
are continuous with respect to an appropriate topology. 
  
Since our considerations are based on C*-algebras, we 
directly obtain, via our reconstruction scheme, 
a Haag-Kastler net of {\em bounded} operators. 
In our case, the technical conditions  
which a reflexion positive functional has to satisfy  
are more natural. It has to be continuous and
regular where the continuity is automatically fulfilled if one 
considers reflexion positive states.  
 
Our reconstruction scheme does also include 
objects, like Wilson loop variables, which are not 
point-like localized objects in a distributional sense. 
This point of view may also be helpful for 
constructing gauge theories. 

Furthermore, one also may start with an abelian C*-algebra like 
the example of Wilson loop variables, given in the introduction. 
Abelian C*-algebras are rather simple objects, namely 
nothing else but continuous functions on a compact 
Hausdorff space. In comparison to the 
construction of reflexion positive functional on 
the tensor algebra $T^\2T_E(S)$, one may hope that it is 
easier to construct reflexion positive functionals for abelian 
C*-algebras. This might simplify the construction of quantum 
field theory models.
\skb

Nevertheless, we also have to mention some drawbacks:

\paragraph{$\ominus$}
Unfortunately, our reconstruction scheme is not a 
complete generalization of the Osterwalder-Schrader reconstruction. 
This is due to that fact, that we have assumed the existence of 
enough operators in $A$ which can be localized on a sharp 
$d-1$-dimensional hyper plane (condition (TZ)). 
Such a condition is not needed within the 
Osterwalder-Schrader framework and there are indeed examples 
of quantum field theories which do not fulfill this 
condition, for instance the generalized free field for 
which the mass distribution is not $L_1$. 

On the other hand, the known interacting models like 
the $P(\phi)_2$, the Yukawa$_2$ as well as the $\phi^4_3$ model 
fulfill the condition (TZ). Thus we think that 
the existence of the time-zero algebras is not such a harmful
requirement.
 
\subsection{Work in progress}

The main aim of our work in progress is concerned with the 
construction of examples for euclidean fields which go 
beyond the free fields. 

It would also be desirable to develop a generalization of 
our reconstruction scheme which also lead directly 
to a Haag-Kastler net but which do not rely on the 
condition (TZ). 

A further open question is concerned with a reconstruction scheme 
for euclidean fields with cutoffs.
The main motivation for such a considerations 
is based on the 
work of J. Magnen, V. Rivasseau, and R. S\'en\'eor \cite{MagRivSen93}
where it is claimed that the Yang-Mills$_4$ exists within a 
finite euclidean volume.


\subsubsection*{{\it Acknowledgment:}}
I am grateful to Prof. Jakob Yngvason for 
supporting this investigation with many ideas.
I am also grateful to Prof. Erhard Seiler and Prof. Jacques Bros 
for many hints and discussions during the workshop at the 
Erwin Schr\"odinger International Institute for Mathematical Physics 
in Vienna (ESI) this autum. 
This investigation is financially supported by the 
Deutsche Forschungsgemeinschaft (DFG) who is also gratefully acknowledged.

\newpage
\begin{appendix}
\section{Analytic properties}
Within this appendix we give a complete proof of 
Lemma \ref{lem21}. We shall use a simplified version of the notation 
introduced in the previous sections by dropping the indices $(e,\eta)$.

Let $(A,\al,\eta)$ be an euclidean field and let $U$ 
be the corresponding strongly continuous representation of the Poincar\'e
group on $\2H=\2H(e,\eta)$ 
which has been constructed by Theorem \ref{thepoin}. 
Furthermore, let $\pi$ be the *-representation of the time-zero algebra 
$B$ on $\2H$.  

For a given tuple $(X,b)\in\8i\6m(d-1)^n\times B^n$, 
we like to study the analytic properties of the function 
\beqa
\Psi_n[X,b]:\7C^{2n}\ni (z,z')\rMapsto 
\prod_{j=1}^nU_{X_j}(z_j)\pi(b_j)U_{X_j}(z'_j)\psi
\eeqa
where $\psi\in D(\Gam,\eta)$ and $\Gam$ is a cone which is contained in 
$\8{Con}(e)$ and we write: 
\beqa
U_X(\zeta):=U(\exp(-\8i\zeta X)) \ \ . 
\eeqa
For this purpose, we introduce some technical definitions.

\bdef
For a generator $X\in\8i\6m(d-1)$, for an operator $b\in B(\2U)$  
and for a cone $\Gam\in\8{Con}(e)$,
we define the regions:
\beqa
I(\Gam,X)&:=&\{s'|\exp(-\8is'X)\Gam\subset e\7R_++\Sgm_e\} 
\vs\vs
J(\Gam,X,b,s')&:=&\{s|\exp(-\8isX)[\exp(-\8is'X)\Gam\cup\2U]
\subset e\7R_++\Sgm_e\}
\vs\vs
G(\Gam|X,b)&:=&\bigcup_{s'\in I(\Gam,X)}[\7R+\8iJ(\Gam,X,b,s')\times
\7R+\8i\{s'\}]
\eeqa
\eef

\bdef
\bdes
\itno 1
Consider a region $\2U$ which is contained in 
$\Sgm_e+e\tau$, $\tau\geq 0$. 
We define the corresponding time-zero algebra by 
$B(\2U):=\al_{e\tau}B(\2U-e\tau)$. 
\itno 2
For a given tuple 
\beqa
(X,b,s,s')\in\8i\6m(d-1)^n\times B(\2U_1)\times\cdots\times B(\2U_n)
\times\7R^{2n}
\eeqa
we define recursively the regions 
\beqa
\Gam_0&:=&\Gam 
\vs\vs
\Gam_1(s_1,s_1')&:=&\8{conv}(\exp(-\8is_1X_1)[\exp(-\8is'_1X_1)\Gam\cup\2U_1])
\vs\vs
\Gam_n(s_1\cdots s_n,s_1'\cdots s_n')&:=&
\8{conv}(\exp(-\8is_nX_n)[\exp(-\8is'_nX_n)\times
\vs
&&
\times\Gam_{n-1}(s_1\cdots s_{n-1},s_1'\cdots s_{n-1}')\cup\2U_n])
\eeqa
\edes
\eef

\bdef
For each $n\in \7N$ we introduce the region: 
\beqa
G_n(\Gam;X,b):=\{(s_1\cdots s_n,s_1'\cdots s_n')|\forall k\leq n:
\Gam_k(s_1\cdots s_k,s_1'\cdots s_k')\subset e\7R_++\Sgm_e\} \ \ .
\eeqa
\eef

\blem\label{lema}
For a given tuple 
\beqa
(X,b)\in\8i\6m(d-1)^n\times B(\2U_1)\times\cdots\times B(\2U_n)
\eeqa
the function $\Psi_n[X,b]$ is holomorphic in $\7R^{2n}+\8iG_n(\Gam;X,b)$.
\elem

\bpr
We prove the statement by induction. 
The vector 
$\psi\in D(\Gam,\eta)$ is contained in the domain of 
$U_{X_1}(\8is_1')$ as long as $s_1'\in I(\Gam,X_1)$.
For a fixed value $s_1'\in I(\Gam,X_1)$ the 
vector $\pi(b_1)U_{X_1}(\8is_1')\psi$ is contained 
in the domain of $U_{X_1}(\8is_1)$ for $s_1\in J(\Gam,X_1,b_1,s_1')$.
This implies that $\Psi_1[X_1,b_1]$ is holomorphic in 
$G(\Gam|X_1,b_1)\supset\7R+\8i G_1(\Gam;X,b)$. 

Suppose $\Psi_{n-1}[X_1\cdots X_{n-1},b_1\cdots b_{n-1}]$ is 
holomorphic in $\7R^{2(n-1)}+\8i G_{n-1}(\Gam;X,b)$. By the same argument as 
above we conclude that for a fixed values 
$(s,s')\in G_{n-1}(\Gam;X,b)$ the function 
\beqa
(z_n,z_n')\rMapsto \Psi_n[X,b]
(\8is,z_n,\8is',z_n')
\eeqa
is holomorphic in
\beqa
G(\Gam_{n-1}(s,s')|X_n,b_n)
\eeqa
and hence it is holomorphic in 
\beqa
\bigcup_{(s,s')\in G_{n-1}(\Gam;X,b)}
\7R^{2(n-1)}+\8i\{(s,s')\}\times G(\Gam_{n-1}(s,s')|X_n,b_n)
\eeqa
which is a region containing $G_n(\Gam;X,b)$.
\epr

\section{Proof of Lemma \ref{lem21}}
For a given euclidean field $(\ul A,\al,\eta)$ we introduce
the following notions:

\bdef
\bdes
\itno 1
We define the subspace 
\beqa
D(\Gam;\eta):=\8p_{(e,\eta)}A(\Gam) \ \mbox{ and } \ \hat D(\Gam;\eta):=
U_{(e,\eta)}(\Lor)D(\Gam;\eta) \ \ .
\eeqa
\itno 2
Let $X\in\8i\6m(d-1)$. For two regions $\Gam_1\subset\Gam$ we define 
\beqa
I(\Gam_1,\Gam;X):=
\{s\in\7R_+|\exp(-\8isX)\Gam_1\subset\Gam\} \ \ .
\eeqa
\itno 3
For a generator $X\in\8i\6m(d-1)$ we define
the region 
\beqa
\2U(s,X):=\exp(-\8isX)\2U
\eeqa
for each $s\in\7R$. 
\itno 4
Given two regions $\2U_1,\2U_2$ in $\7R^d$, we define 
\beqa
G_e(X_1,X_2;\2U_1,\2U_2;I)&:=&\biggl\{(z_1,z_2)\in(\7R\times\8iI)^2\biggm 
|\forall \1x_j\in\2U_j:
\vs\vs
&&e \ \8{Im}(\exp(z_1X_1)\1x_1-\exp(z_2X_2)\1x_2)\in \7R_+ \biggr\} 
\vs\vs
G^g_e(X_1,X_2;\2U_1,\2U_2;I)&:=&\biggl\{(z_1,z_2)\in(\7R\times\8iI)^2\biggm 
|\forall \1x_j\in\2U_j:
\vs\vs
&&e \ \8{Im}[g(\exp(z_1X_1)\1x_1-\exp(z_2X_2)\1x_2)]\in \7R_+ \biggr\} 
\eeqa
where $g\in\8P_+(\7C)$ is a complex Poincar\' e transformation. 
\edes
\eef

\blem\label{lemhola}
Let $\Gam_1,\Gam\in\8{Con}(e)$ be two conic regions 
such that $g\Gam_1\subset\Gam$ is a proper inclusion. 
Then there exists an interval $I$ such that 
for each $b_1\in B(\2U_1),b_2\in B(\2U_2)$ 
and for each $\psi_1,\psi_2\in \hat D(\Gam_1;\eta)$ the function 
\beqa
F^{(\psi_1,\psi_2)}_{(X_1,X_2,b_1,b_2)}:(z_1,z_2)\rMapsto 
\<\psi_1,\Phi_{X_1}(z_1,b_1)\Phi_{X_2}(z_2,b_2)\psi_2\>
\eeqa
is holomorphic in $G_e(X_1,X_2;\2U_1,\2U_2;I)$.
\elem

\bpr
First we obtain by an application of Lemma \ref{lema},  
that for each $\psi_1\in\2H(e,\eta)$ 
and for each $\psi\in D(\Gam,\eta)$, the function 
\beqa
(z,\zeta)\rMapsto
\<\psi_1,\Phi_{X_2}(z,b_2)U_X(\zeta)\psi_2\>
\eeqa
is holomorphic for $\8{Im}\zeta\in I(\Gam_1,\Gam;X)$ and 
$\8{Im}z\in I(\Gam;X_2)$. 
for $X\in\8i\6m(d-1)$. The holomorphy is due to the fact that 
$U$ is a strongly continuous representation of the 
Poincar\' e group and that $D(\Gam;\eta)$ consists of 
analytic vectors for the boost generators.  

For a fixed values $s'\in I(\Gam_1,\Gam;X)$ and $s\in I(\Gam;X_2)$, we have 
\beqa
\Phi_{X_2}(\8is,b_2)U_{(e,\eta,X)}(-\8is')\psi_2\in 
D(\hat\Gam;\eta)
\eeqa
for each region $\hat\Gam\subset e\7R_++\Sgm_e$ 
which contains $\Gam\cup\2U_2(s,X_2)$.  
\skb

Now, for a given point $(z,\8is)\in G_e(X_1,X_2;\2U_1,\2U_2;I)$
there exists a conic region $\Gam(z,\8is)\in\8{Con}(e)$ with 
$\Gam(z,\8is)\supset\Gam\cup\2U_2(s,X_2)$ such that  
$D(\Gam(z,\8is);\eta)$ is contained in the domain of 
$\Phi_{X_1}(z,b_2)$. Furthermore, for a
given interval $I$, the cone $\Gam$ can be 
chosen to be small enough such that this holds for 
each $(z,\8is)$ with $\8{Im}z,s\in I$.
Since $\Gam_1$ is $O(d-1)$-invariant, the result follows.
\epr

Let $\2V\supset\Lor$ be a neighborhood of the identity in $\8P_+(\7C)$.
We may choose a cone $C(\Gam,\2V)\in\8{Con}(e)$
such that 
\beqa
g C(\Gam,\2V)\subset \Gam \ \ . 
\eeqa
for each $g\in \8E(d)\cap\2V$. Note that the representation 
$U$ can be extended to $\2V$ by unbounded operators 
with domain $\hat D(\Gam_1,\eta)$ where $\Gam_1\subset C(\Gam,\2V)$.

In order to finish the proof of Lemma \ref{lem21}, we show the 
following statement:

\blem\label{lemholb}  
Let $\2U_1,\2U_2$ be two bounded disjoint regions
and let $\Gam_1\in\8{Con}(e)$ such that $\Gam_1\subset C(\Gam,\2V)$ 
is a proper inclusion.  
Then the function  $F^{(\psi_1,\psi_2)}_{(X_1,X_2,b_1,b_2)}$
has an extension $\hat F^{(\psi_1,\psi_2)}_{(X_1,X_2,b_1,b_2)}$
which is holomorphic in 
\beqa
G(\2V;X_1,X_2;\2U_1,\2U_2;I):=
\bigcup_{g\in\2V} G_e^g(X_1,X_2;\2U_1,\2U_2;I)
\eeqa
for each $\psi_1,\psi_2\in \hat D(\Gam_1;\eta)$. 
\elem

\bpr
For a given neighborhood $\2V\supset \Lor$ of the unit element in $\8P_+(\7C)$
and for a given cone $\Gam\in\8{Con}(e)$, there 
exists $\eps>0$ such that  $g\2U_2+\eps e\subset \Gam$.
We easily observe that the substitution  
\beqa
\psi_j'&:=&T(\eps)U(g)\psi_j
\vs\vs
X_j'&:=& \exp(-\8i\eps H)gX_jg^{-1}\exp(\8i\eps H)
\eeqa
yields 
\beqa
F^{(\psi_1',\psi_2')}_{(X_1',X_2',b_1,b_2)}(z_1,z_2)=
F^{(\psi_1,\psi_2)}_{(X_1,X_2,b_1,b_2)}(z_1,z_2)
\eeqa
for each $(z_1,z_2)\in G_e(X_1,X_2;\2U_1,\2U_2;I)$ where 
$H$ is the generator of translations in $e$-direction.
According to Lemma \ref{lemhola}, 
the function $F^{(\psi_1',\psi_2')}_{(X_1',X_2',b_1,b_2)}$ is 
holomorphic in $G^g_e(X_1,X_2;\2U_1,\2U_2;I)$ 
which implies the result.
\epr
\end{appendix}
 
\newpage


\end{document}